\documentclass[12pt,twoside]{article}
\usepackage{epsfig}

\newcommand{\beq}{\begin{equation}}
\newcommand{\eeq}{\end{equation}}
\newcommand{\bea}{\begin{eqnarray}}
\newcommand{\eea}{\end{eqnarray}}

\newcommand{\nrad}{\ensuremath{n_{\mathrm{rad}}}}

\newcommand{\gr}{\ensuremath{\tilde{G}}}
\newcommand{\mgr}{\ensuremath{m_{\gr}}}
\newcommand{\ngr}{\ensuremath{n_{\gr}}}
\newcommand{\Ygr}{\ensuremath{Y_{\gr}}}
\newcommand{\Ogr}{\ensuremath{\Om_{\gr}}}
\newcommand{\ONSP}{\ensuremath{\Om_{\mathrm{NSP}}}}
\newcommand{\gl}{\ensuremath{\tilde{g}}}
\newcommand{\mgl}{\ensuremath{m_{\gl}}}
\newcommand{\sq}{\ensuremath{\tilde{q}}}
\newcommand{\msq}{\ensuremath{m_{\sq}}}

\newcommand{\wfnx}{\ensuremath{\psi_\mu(x)}}

\newcommand{\Pmn}{\ensuremath{\Pi_{\mu\nu}(P)}}

\newcommand{\slK}{\ensuremath{/ \! \! \! \! K}}
\newcommand{\slP}{\ensuremath{/ \! \! \! \! P}}
\newcommand{\slQ}{\ensuremath{/ \! \! \! \! Q}}
\newcommand{\slv}{\ensuremath{/ \! \! \! \! v}}

\newcommand{\slder}{\ensuremath{/ \! \! \! \partial}}
\newcommand{\pk}{\ensuremath{\mathbf{p}\mathbf{k}}}

\newcommand{\Gs}{\ensuremath{\G_{\gr}^{\mbox{\tiny soft}}}}

\newcommand{\factor}{\ensuremath{\left(1 + \frac{\mgl^2}{3\mgr^2} \right)}}
\newcommand{\ttt}{2\rightarrow 2}

\newcommand{\al}{\ensuremath{\alpha}}

\newcommand{\g}{\ensuremath{\gamma}}
\newcommand{\G}{\ensuremath{\Gamma}}

\newcommand{\D}{\ensuremath{\Delta}}
\newcommand{\e}{\ensuremath{\varepsilon}}

\newcommand{\la}{\ensuremath{\lambda}}

\newcommand{\s}{\ensuremath{\sigma}}
\newcommand{\Si}{\ensuremath{\Sigma}}

\newcommand{\om}{\ensuremath{\omega}}
\newcommand{\Om}{\ensuremath{\Omega}}

\newcommand{\Lc}{\ensuremath{\mathcal{L}}}
\newcommand{\Mc}{\ensuremath{\mathcal{M}}}

\newcommand\ap[3]{Annals~Phys.~{\bf #1} (19{#2}) #3}
\newcommand\app[3]{Astropart.~Phys.~{\bf #1} (19{#2}) #3}
\newcommand\np[3]{Nucl.~Phys.~{\bf B#1} (19{#2}) #3}
\newcommand\pa[3]{Physica A {\bf #1} (19{#2}) #3}
\newcommand\pl[3]{Phys.~Lett.~{\bf B #1} (19{#2}) #3}

\newcommand\pr[3]{Phys.~Rev.~{\bf #1} (19{#2}) #3}
\newcommand\prd[3]{Phys.~Rev.~{\bf D#1} (19{#2}) #3}

\newcommand\prl[3]{Phys.~Rev.~Lett.~{\bf #1} (19{#2}) #3}

\newcommand\ptp[3]{Progr.~Theor.~Phys.~{\bf {#1}} (19{#2}) #3}
\newcommand\sjnp[3]{Sov.~J.~Nucl.~Phys.~{\bf #1} (19{#2}) #3}
\newcommand\spj[3]{Sov.~Phys.~JETP~{\bf #1} (19{#2}) #3}

\begin{document}
\date{\mbox{ }}
\title{{\normalsize DESY 00-167\hfill\mbox{}\\
April 2007 (revised)\hfill\mbox{}}\\
\vspace{2cm} \textbf{Thermal Production of Gravitinos}\\
[8mm]}
\author{M.~Bolz, A.~Brandenburg, W.~Buchm\"uller \\
\textit{Deutsches Elektronen-Synchrotron DESY, Hamburg, Germany}}
\maketitle
\thispagestyle{empty}
\begin{abstract}
\noindent
We evaluate the gravitino production rate in supersymmetric QCD at high
temperature to leading order in the gauge coupling. The result, which is 
obtained by using the resummed gluon propagator, depends logarithmically on
the gluon plasma mass. As a byproduct, a new result for the axion
production rate in a QED plasma is obtained. The implicatons for the
cosmological dark matter problem are briefly discussed, in particular
the intriguing possibility that gravitinos are the dominant part of
cold dark matter.
\end{abstract}

\newpage
\section{Introduction}
\label{intro}
\ 
\vskip -0.5cm
Supersymmetric theories, which contain the standard model of particle physics
and gravity, predict the existence of the gravitino \cite{Des76}, 
a spin-${3\over 2}$
particle which aquires a mass from the spontaneous breaking of supersymmetry.
Since the couplings of the gravitino with ordinary matter are strongly
constrained by local supersymmetry, processes involving gravitinos allow 
stringent tests of the theory.   

It was realized long ago that standard cosmology requires gravitinos to
be either very light, $\mgr<1$\,keV \cite{Pag82}, or 
very heavy, $\mgr>10$\,TeV \cite{Wei82}. These constraints 
are relaxed if the standard cosmology
is extended to include an inflationary phase \cite{Khl84,Ell84}.
The cosmologically relevant gravitino abundance is then created in the 
reheating phase after inflation in which a reheating temperature $T_R$ is
reached. Gravitinos are dominantly produced by inelastic $\ttt$ scattering 
processes of particles from the thermal bath. The gravitino abundance is 
essentially linear in the reheating temperature $T_R$. 

The gravitino production rate depends on $\mgl / \mgr$, the ratio
of gluino and gravitino masses. The ten $\ttt$ gravitino production processes 
were considered in \cite{Ell84} for $\mgl \ll \mgr$. The case $\mgl \gg \mgr$,
where the goldstino contribution dominates, was considered in \cite{Mor93}.
Four of the ten production processes are logarithmically singular due to the 
exchange of massless gluons. As a first step this 
singularity can be regularized
by introducing either a gluon mass or an angular cutoff \cite{Ell84}. The
complete result for the logarithmically singular part of the production rate
was obtained in \cite{Bol98}. The finite part depends on the cutoff procedure.

To leading order in the gauge coupling the correct finite result for the
gravitino production rate can be obtained by means of a hard thermal loop
resummation. This has been shown by Braaten and Yuan in the case of axion
production in a QED plasma \cite{Bra91}. The production rate is defined
by means of the imaginary part of the thermal axion self-energy \cite{Wel83}. 
The different contributions are split into  parts with soft and hard 
loop momenta by means of a momentum cutoff. For the soft part a resummed 
photon propagator is used, and the logarithmic singularity, which appears
at leading order, is regularized by the plasma mass of the photon. The
hard part is obtained by computing the $\ttt$ scattering processes with
momentum cutoff. In the sum of both contributions the cutoff dependence
cancels and the finite part of the production rate remains. For the
gravitino production rate the soft part has been considered in \cite{Ell96}
and the expected logarithm of the gluon plasma mass has been obtained.

Constraints from primordial nucleosynthesis imply an upper bound on the
gravitino number density which subsequently yields an upper bound on the 
allowed reheating temperature $T_R$ after inflation \cite{Ell92}-\cite{Asa00}. 
Typical values for $T_R$ range from $10^7 - 10^{10}$~GeV, although 
considerably larger temperatures are acceptable in some cases \cite{AY00}.
In models of baryogenesis where the cosmological baryon asymmetry
is generated in heavy Majorana neutrino decays \cite{Fuk86},
temperatures $T_R\simeq 10^8 - 10^{10}$~GeV are of particular 
interest \cite{Buc00}. 
Further, it is intriguing that for such temperatures gravitinos with mass of the electroweak scale, i.e. $\mgr \sim 100$~GeV can be the dominant component
of cold dark matter \cite{Bol98}. In all these considerations the
thermal gravitino production rate plays a crucial role. In this paper  we 
therefore calculate this rate to leading order in the gauge coupling, 
extending a previous 
result \cite{Bol98} and following the procedure of Braaten and Yuan 
\cite{Bra91}.

The paper is organized as follows. In section~2 we summarize some properties
of gravitinos and their interactions which are needed in the following. 
In order to illustrate how the hard thermal loop resummation is incorporated
we first discuss the axion case in section~3. The most important intermediate
steps and the final result for the gravitino production rate are given in
section~4. Using the new results the discussion in \cite{Bol98} on gravitinos
as cold dark matter is updated in section~5, which is followed by an outlook
in section~6. The calculation of the hard momentum contribution to the
production rates is technically rather involved. We therefore give the
relevant details in the appendices. 

\section{Gravitino interactions}
\label{chap:spinth}
\ 
\vskip -0.5cm
In the following we briefly summarize some properties of gravitinos
which we shall need in the following sections. More detailed discussions and
references can be found in \cite{Wes92,Mor95,Wei00}.

Gravitinos are spin-$3/2$ particles whose properties are given by the
lagrangian for the vector-spinor field $\psi^\al_\mu(x)$, 
\begin{equation}
\Lc = 
-\frac{1}{2}\e^{\mu\nu\rho\s}\overline{\psi}_\mu\g_5\g_\nu\partial_\rho\psi_\s
-\frac{1}{4}m_{\gr}\overline{\psi}_\mu[\g^\mu,\g^\nu]\psi_\nu
-{1\over 2 M}\overline{\psi}_\mu S^{\mu}\;.
\label{eq:lagrangian}
\end{equation}
Here $m_{\gr}$ is the gravitino mass, $M=(8\pi G_N)^{-1/2}$ is the Planck mass
and $S_{\mu}$ is the supercurrent corresponding to supersymmetry 
transformations. $\psi_\mu$ and $S_{\mu}$ are Majorana fields, so that
$\overline{\psi}_\mu S^{\mu} =\overline{S}_\mu \psi^{\mu}$. 

Free gravitinos satisfy the Rarita-Schwinger equation,
\begin{equation}
-\frac{1}{2}\e^{\mu\nu\rho\s}\g_5\g_\nu\partial_\rho\psi_\s
-\frac{1}{4} m_{\gr}[\g^\mu,\g^\nu]\psi_\nu = 0 \;,
\end{equation}
which, using 
\beq
\g^\mu \wfnx  =  0 \;, \quad \partial^\mu\wfnx = 0\;,
\eeq
reduces to the Dirac equation
\beq
(i\slder-m_{\gr}) \wfnx  =  0 \;.   
\eeq

Consider as matter sector first a non-abelian supersymmetric gauge theory
with lagrangian
\beq
\Lc = -{1\over 4} F^a_{\mu\nu}F^{a\mu\nu}
      +{1\over 2} \overline{\lambda^a}\left(
        i\g^\mu D_{ab\mu} - m_{\gl} \delta_{ab}\right)\lambda^b
\eeq
for the vector boson $A^a_\mu$ and the gluino $\lambda^a$. Supersymmetry
is explicitly broken by the gluino mass term. Hence, the supercurrent is
not conserved,
\bea\label{pcsc}   
\partial_\mu S^\mu &=& \partial_\mu 
{i\over 4}[\g^\nu,\g^\rho]\g^\mu \lambda^a F^a_{\nu\rho} \nonumber\\
&=& m_{\gl} {1\over 4}[\g^\nu,\g^\rho]\lambda^a F^a_{\nu\rho} \nonumber\\
&=& m_{\gl} S\;.
\eea

The calculation of the gravitino production rate in section~4 will involve
squared matrix elements which are summed over all four gravitino polarizations.
The corresponding polarization tensor for a gravitino with momentum $P$ reads
\bea
\Pmn &=& \sum_l \psi^l_\mu(P)\overline{\psi}^l_\nu(P) \nonumber\\
&=&-(\slP+m_{\gr})\left(g_{\mu\nu} -\frac{P_\mu P_\nu}{m_{\gr}^2}\right) 
-\frac{1}{3} \left(\g^\mu + \frac{P_\mu}{m_{\gr}}\right)
(\slP-m_{\gr})
\left(\g^\nu + \frac{P_\nu}{m_{\gr}}\right)\;.
\label{eq:Pmn}
\eea
Since $\psi_\mu(x)$ is a solution of the Rarita-Schwinger equation one has
for the polarization tensor 
\beq
\g^\mu \Pmn  =  0\;, \quad P^\mu \Pmn = 0 \;, \nonumber
\eeq
\beq
(\slP-m) \Pmn  =  0 \;.   
\eeq

We shall be interested in the production of gravitinos at energies much larger
than the gravitino mass. In this case the polarization tensor simplifies to
\beq\label{decomp}
\Pmn \simeq - \slP g_{\mu\nu} 
       + {2\over 3} \slP {P_\mu P_\nu\over m_{\gr}^2}  \;,
\eeq
where we have used $\g^\mu S_\mu =0$.
Clearly, the first term corresponds to the sum over the helicity 
$\pm {3\over 2}$ states and the second term represents the sum over the
helicity $\pm {1\over 2}$ states which represent the goldstino part of
the gravitino. Eq.~(\ref{decomp}) is the basis for the familiar substitution
rule $\psi_\mu \rightarrow \sqrt{{2\over 3}}{1\over m_{\gr}}\partial_\mu \psi$
which is used to obtain the effective lagrangian describing the interaction
of goldstinos with matter \cite{Fay79,Cla96,Lee99}.

The gravitino production rate at finite temperature can be expressed in terms
of the imaginary part of the gravitino self-energy \cite{Wel83} which takes
the form,
\bea\label{decomp1}
\Sigma(P) &=& \mbox{tr}{\left[\Pmn \Si_{\gr}^{\nu\mu}(P)\right]} \nonumber\\
&\propto&  {1\over M^2} \mbox{tr}{\left[\Pmn S^{\nu}(P)\ldots
\overline{S}^\mu(P)\right]}\nonumber\\ 
&\propto& {1\over M^2} \mbox{tr}\left[
(- \slP) S^\mu(P)\ldots \overline{S}_\mu(P)\right] \nonumber\\
&& \hspace{1cm} +  {2 m_{\gl}^2\over 3 m_{\gr}^2 M^2} \mbox{tr}\left[
\slP S(P) \ldots\overline{S}(P)\right] \;.
\eea
Here we have used eq.~(\ref{pcsc}) for the divergence of the supercurrent. 
Note, that $F=\sqrt{3}m_{\gr}M$ is the scale of spontaneous supersymmetry 
breaking, which gives the strength of the goldstino coupling to the 
supercurrent. The dots denote the sum over the contributions to the
self-energy in the loop expansion. In $S^\mu(P)={i\over 4}[\g^\nu,\g^\rho]
\g^\mu \lambda^a(P_1)F_{\nu\rho}(K_1)$ and $S(P)={1\over 4}[\g^\nu,\g^\rho]
\lambda^a(P_1)F_{\nu\rho}(K_1)$, with $P=P_1+K_1$, $\lambda^a(P_1)$ represents
one end of an internal gluino line and $F_{\nu\rho}(K_1)$ stands for the end
of one or two internal gluon lines. For $m_{\gr} \ll m_{\gl}$ the goldstino 
part dominates the gravitino production cross section.

Eq.~(\ref{decomp1}) is useful to derive relations between the
helicity $\pm {3\over 2}$ and the helicity $\pm {1\over 2}$ contibutions to
the self-energy. As shown in Appendix~A, one obtains to two-loop order
\beq\label{factor}
\Sigma(P) \propto \frac{g^2}{M^2}\factor\;.
\eeq
We shall exploit this fact to perform the hard thermal loop resummation,
which is necessary because of the infrared divergences, just for the helicity 
$\pm {1\over 2}$ part of the rate.

The full supercurrent also involves quarks and squarks in addition to 
gluons and gluinos. The divergence of the additional part of the
supercurrent that involves the cubic gravitino-quark-squark coupling 
is again proportional to the parameter of 
supersymmetry breaking, i.e. $m^2_{\tilde{q}}$, the squark mass squared. 
The corresponding goldstino contribution to the production rate is then
proportional to $m^4_{\tilde{q}}$, 
which is suppressed at high energies compared to the gluino contribution
for dimensional reasons.
 
\section{Axion production}
\label{chap:selfenergy}
\ 
\vskip -0.5cm
Let us now consider the thermal production of axions in a relativistic
QED plasma of electrons and photons. According to the procedure outlined
in the introduction the thermal production rate can be obtained as sum
of two terms, a soft momentum contribution which is extracted from the
axion self-energy evaluated with a resummed photon propagator and a hard 
momentum contribution which is computed from the $\ttt$ processes.

\begin{figure}[h]
  \begin{center}
  \epsfig{file=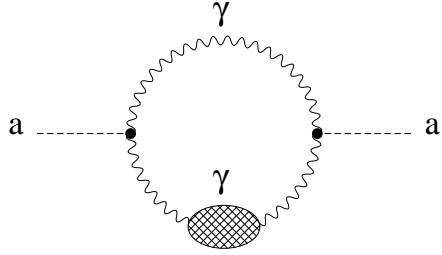,width=14em}
  \caption{\it Axion self energy; the blob denotes the
    resummed photon propagator.}
  \label{fig:axionloop}
  \end{center} 
\end{figure} 

The axion-photon interaction is described by the effective lagrangian
\beq
{\cal L} = - {1\over 4f} a F_{\mu\nu}\widetilde{F}^{\mu\nu}\;,
\eeq 
where $f$ is the axion decay constant. The axion self-energy $\Pi_a(P)$ 
(cf.~fig.~\ref{fig:axionloop})
can be evaluated in the imaginary-time formalism for external momentum
$P=(p_0,{\bf p})$, with $p_0=i2\pi n T$ and $p=|{\bf p}|$. In covariant gauge
the resummed photon propagator has the form \cite{gluonprop,Pis89}
\beq
  \label{eq:effgluonprop}
  i\D_{\mu\nu}(K) = i\left( 
  A_{\mu\nu} \D_T + B_{\mu\nu} \D_L + C_{\mu\nu} \xi  \right)\;,
\eeq
with the tensors 
\bea
  \label{eq:tensors}
  A_{\mu\nu} & = & -g_{\mu\nu} -\frac{1}{k^2}\left[
K^2 v_\mu v_\nu - K\cdot v (K_\mu v_\nu + K_\nu v_\mu) + K_\mu K_\nu 
\right]\;, \nonumber \\
  B_{\mu\nu} & = & v_\mu v_\nu - 
\frac{K\cdot v}{K^2}(K_\mu v_\nu + K_\nu v_\mu) + \left(
\frac{K\cdot v}{K^2}\right)^2 K_\mu K_\nu\;, \nonumber \\
  C_{\mu\nu} & = & \frac{K_\mu K_\nu}{\left(K^2\right)^2}\;,
\eea
and the transverse and longitudinal propagators
\bea
  \label{eq:ltprop}
  \D_T(k_0,k) & = & \frac{1}{k_0^2-k^2 - \Pi_T(k_0,k)}\;,\nonumber \\
  \D_L(k_0,k) & = & \frac{1}{k^2 - \Pi_L(k_0,k)}\;.
\eea
Here $K=(k_0,{\bf k})$ with $k_0=i2\pi n T$ and $k=|{\bf k}|$,
$\xi$ is a gauge-fixing parameter, $v$ is the velocity of the thermal 
bath and $\Pi_{T/L}$ are the transverse and longitudinal self-energies 
of the photon.
The corresponding propagators $\D_{T/L}$ have the spectral representation
\beq
  \D_{T/L}(k_0,k)  =  \int_{-\infty}^\infty d\om 
   {1\over k_0-\om}  \rho_{L/T}(\om,k)\;.
\eeq
For $|\om|<k$ the spectral densities $\rho_{L/T}$ are given by \cite{Pis89},
\bea
  \label{eq:gluonspectra}
  \rho_T(\om,k) & = & 
  \frac{3}{4m_\g^2}\frac{x}{(1-x^2)(A_T(x)^2+(z+B_T(x))^2)}\;,
  \nonumber\\
  \rho_L(\om,k) & = & \frac{3}{4m_\g^2} \frac{2 x}{A_L(x)^2+(z+B_L(x))^2}\;,
\eea
where $m_\g = eT/3$ is the plasmon mass of the photon, $x = \om/k$, 
$z = k^2/m_\g^2$ and
\bea
  \label{eq:ablt}
  A_T(x) = \frac{3}{4}\pi x, & & B_T(x) = 
  \frac{3}{4}\left( 2\frac{x^2}{1-x^2} + x
  \ln{\frac{1+x}{1-x}}\right)\;,\nonumber\\
  A_L(x) = \frac{3}{2}\pi x, & & B_L(x) = 
  \frac{3}{2}\left( 2 - x \ln{\frac{1+x}{1-x}}\right)\;.
\eea

The contribution to the axion production rate from soft virtual photons
is obtained by analytically continuing the axion self energy function
$\Pi_a(P)$ from the discrete imaginary
value $p_0$ to the continuous real value $E=p$ \cite{Bra91},
\bea
  \label{eq:gammasoft1}
\G_a^{\rm soft}(E) &=& 
\left.-\frac{\mbox{Im}{\Pi_a(E+i\epsilon,p)}}{E}\right|_{k<k_{\rm cut}}
\nonumber \\
&=& \frac{T}{8\pi f^2} \int_0^{k_{\rm cut}} dk k^3 \int_{-k}^k {d\om\over \om} 
\left[\rho_L(\om,k)\left(1-{\om^2\over k^2}\right) 
+ \rho_T(\om,k)\left(1-{\om^2\over k^2}\right)^2 \right] .
\eea
The axion production rate depends logarithmically on $k_{\rm cut}$. The
corresponding coefficient can be obtained analytically. The remaining constant
has to be evaluated numerically. 
This yields the result, first obtained in \cite{Bra91},
\beq
  \label{eq:gammasoft2}
  \G_a^{\rm soft}(E) = {3 m_\g^2 T\over 16\pi f^2}
  \left[\ln{{k^2_{\rm cut}\over m^2_\g}}-1.379\right]\;.
\eeq
The corresponding collision term in the Boltzmann equation is
\bea\label{csoft}
C_a^{\rm soft}(T) &=& \int {d^3p\over (2\pi)^3} n_B(E) 
\G_a^{\rm soft}(E) \nonumber\\
&=&\frac{e^2\zeta(3)T^6}{24\pi^3f^2}
\left[\ln\left(\frac{k_{\rm cut}}{m_\g}\right)-0.689\right]\;,
\eea
where
\beq
n_B(E) = {1\over \exp{(E/T)}-1}
\eeq
is the Bose-Einstein distribution.
\begin{figure}[h]
  \begin{center}
  \epsfig{file=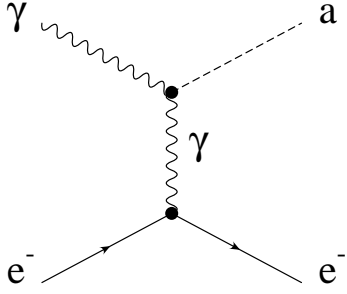,width=12em}
  \caption{\it Axion production in electron-photon scattering.}
  \label{fig:axion22}
  \end{center} 
\end{figure} 

The dependence of $\G_a^{\rm soft}(E)$ on the cutoff $k_{\rm cut}$ is 
cancelled by the contribution from hard virtual photons to the self-energy. 
This part of the axion production rate can be obtained directly from the 
processes $e^{\pm}\g \rightarrow e^{\pm} a$ (cf.~fig.~\ref{fig:axion22}) 
\cite{Wel83},
\bea
\label{eq:harddef}
n_B(E)\G_a^{\rm hard}(E) 
&=& 2\int {d\Om_p\over 4\pi}{1\over 2E}\int\left[\prod_{i=1}^3
\frac{d^3p_i}{(2\pi)^3 2E_i}\right](2\pi)^4\delta^4(P_1+P_2-P-P_3)\nonumber \\
&& \ \ \ n_F(E_1)n_B(E_2)(1 - n_F(E_3))
|M|^2\Theta(|{\bf p}_1-{\bf p}_3|-k_{\rm cut})\;.
\eea
Here we have used rotational invariance by averaging over the directions of
the axion momentum; $|M|^2$ is the photon-axion matrix element squared for
 $e^-(P_1) \g(P_2) \rightarrow e^-(P_3) a(P)$,
\beq\label{eq:axionmat}
|M|^2 = {e^2\over f^2}\left(-\frac{2s^2}{t}-2s-t\right)\;,
\eeq
where $s=(P_1+P_2)^2,\ t=(P_1-P_3)^2$, 
and $n_{F}(E)$ is the Fermi-Dirac distribution, 
\beq
n_F(E)=\frac{1}{\exp(E/T)+1}\;.
\eeq
The phase space integration has to be carried out under the constraint on
the virtual photon momentum $k\equiv
|{\bf p}_1-{\bf p}_3| > k_{\rm cut}$. For the angular 
integrations it turns out to be convenient to define all momenta with
respect to ${\bf k}$. Some details of this calculation are given in 
appendix~B. One finally obtains,
\bea
\label{eq:phiint}
n_B(E)\G_a^{\rm hard}(E) 
&=& \frac{3 e^2}{2^8\pi^3 f^2} {1\over E^2} \int dE_1 dE_3 
dk n_F(E_1)n_B(E_2)(1 - n_F(E_3))
\nonumber \\&&\hspace{1.5cm}\times 
\left[(E_1-E_3)^2-k^2\right]
\left(-1+\frac{2}{3}\frac{E_1^2+E_3^2+2EE_2}{k^2}
\right.\nonumber \\
&&\hspace{1.5cm}
\left.-\frac{(E_3+E_1)^2(E+E_2)^2}{k^4}\right)\ \Om\;,
\eea
where $E_2=E+E_3-E_1$ and the integrations are restricted by $\Om$, 
\bea\label{eq:omega}
\Omega &=& \Theta(k-k_{\rm cut})\Theta(k-|E_1-E_3|)\nonumber \\
&& \Theta(E_1+E_3- k)\Theta(2E+E_3-E_1- k)\nonumber\\
&&\Theta(E_1)\Theta(E_3)\Theta(E+E_3-E_1)\;.
\eea
After performing the $k$-integration one is left with several domains
for the $E_1$- and $E_3$-integrations. The logarithmic dependence on 
$k_{\rm cut}$ can be extracted by means of a partial integration in $E_1$.
In the remaining part of the integral $k_{\rm cut}$ can be set equal to zero.
The final result reads
\bea\label{chard}
\G_a^{\rm hard}(E)&=&\frac{e^2}{16\pi^3 f^2}\Bigg\{{2\pi^2\over 3}T^3\left(
\ln\left(\frac{2T}{k_{\rm cut}}\right)+\frac{17}{6}-\gamma+\frac{\zeta'(2)}
{\zeta(2)}\right)  \nonumber \\
&&+  \left(e^{E/T}-1\right)
\int_0^\infty dE_3 \left[1-n_F(E_3)\right]\int_0^{E+E_3}dE_1 \nonumber \\
&&\hspace{.5cm}\times
\ln\left(\frac{|E_1-E_3|}{E_3}\right)n_F(E_1)n_B(E_2)\nonumber \\ 
&&\hspace{.5cm}\times
\Bigg[\Theta(E-E_1)\frac{E_2^2}{E^2}\left(2E_1-2\frac{E_1^2}{E_2}
+\left(E_1^2-\frac{E_3^2E^2}{E_2^2}\right)\frac{n_F(E_1)+n_B(E_2)}{T}\right)
\nonumber \\
&&\hspace{0.8cm} -\Theta(E_1-E_3)\frac{E_2^2}{E^2}
\left(2E_1-2\frac{E_1^2+E_3^2}{E_2}
+(E_1^2+E_3^2)\frac{n_F(E_1)+n_B(E_2)}{T}\right)\nonumber \\ 
&&\hspace{0.8cm} +
\Theta(E_3-E_1)\left(2E_1+(E_1^2+E_3^2)\frac{n_F(E_1)+n_B(E_2)}{T}\right)
\Bigg]\Bigg\}\;.
\eea
This result agrees with the one obtained in \cite{Bra91} except for the
first expression $\propto \Theta(E-E_1)$ in the double integral. Integration
over the axion energy $E$ yields for the collision term in the Boltzmann
equation
\bea
C_a^{\rm hard}(T) &=& \int {d^3p\over (2\pi)^3} n_B(E) 
\Gamma_a^{\rm hard}(E) \nonumber\\
&=& \frac{e^2\zeta(3)T^6}{24\pi^3f^2}\left[
\ln\left(\frac{2T}{k_{\rm cut}}\right)+\frac{17}{6}-\gamma+\frac{\zeta'(2)}
{\zeta(2)}-1.280\right]\;.
\eea
The numerical constant is about 20\% smaller than the one obtained from the
axion rate given in \cite{Bra91}.

Consistency requires that the dependence on the cutoff $k_{\rm cut}$ cancels
in the total production rate. Comparison of eqs.~(\ref{csoft}) and 
(\ref{chard}) shows that this is indeed the case. The result for the total 
axion collision term reads 
\bea
C_a(T) &=& C_a^{\rm soft}(T) + C_a^{\rm hard}(T) \nonumber\\
&=& \frac{e^2\zeta(3)T^6}{48\pi^3f^2}
\left[\ln\left(\frac{T^2}{m_{\gamma}^2}\right) + 0.8194\right]\;.
\eea

\section{Gravitino Production}
\ 
\vskip -0.5cm
The rate for the thermal production of gravitinos can be calculated in
complete analogy to the axion production rate. It
is dominated by QCD processes since the strong coupling is 
considerably larger than the electroweak couplings. The contribution due to
soft virtual gluons can again be extracted from the gravitino self-energy
with a resummed gluon propagator to which the contribution from hard $\ttt$ 
processes has to be added.

Properties and interactions of the gravitino have been 
discussed in section~2.
For supersymmetric QCD with gluons, gluinos, quarks and squarks one obtains 
\cite{Wes92,Mor95,Wei00},
\beq
  \Lc = -\frac{i}{\sqrt{2}M}
  \left[(D^\ast_\mu \phi^\ast) \overline{\psi_\nu} \g^\mu \g^\nu
  P_L \chi - (D_\mu \phi) \overline{\chi} P_R \g^\nu \g^\mu \psi_\nu\right] -
  \frac{i}{8 M}\overline{\psi_\mu} \left[ \g^\nu, \g^\rho \right]
  \g^\mu\lambda^a F^a_{\nu\rho}.
  \label{eq:lag}
\eeq
Here $\chi$ denotes a left-handed quark or antiquark and $\phi$ the 
corresponding squark. For light gravitinos one can use a simpler effective 
lagrangian \cite{Fay79,Cla96,Lee99}. The corresponding goldstino-gluon-gluino
coupling can be read off from eqs.~(\ref{pcsc}) and (\ref{decomp1})
\beq\label{leff}
  \Lc^{\mathrm{eff}} =  -
  \frac{\mgl}{2\sqrt{6}M\mgr}\overline{\psi}\left[\g^\mu,\g^\nu\right]
  \la^a F^a_{\mu\nu} +\ldots \;.
\eeq
Here $\psi$ is the goldstino, the spin-$1/2$ component of the gravitino. 
The effective theory contains the same vertices as the full theory, except for the gravitino-quark-squark-gluon vertex. Instead, there is a new four 
particle vertex, 
the gravitino-gluino-squark-squark vertex \cite{Lee99}. 
All vertices are proportional to supersymmetry breaking mass terms, i.e., 
$\msq^2$ and $\mgl$. At high energies and temperatures, with $\msq,\mgl \ll T$,
contributions involving the cubic  goldstino-quark-squark coupling are
suppressed by $\msq^2/T^2$ relative to the gluino 
contribution because of the higher mass dimension of the coupling.

\begin{figure}[h]
  \begin{center}
  \epsfig{file=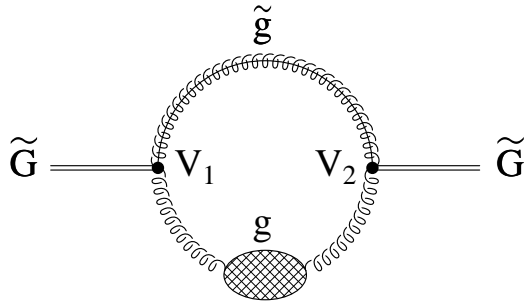,width=18em}
  \caption{\it Gluon-gluino loop diagram, the leading contribution to
    the imaginary part of the gravitino self energy. The blob denotes
    a resummed gluon propagator.}
  \label{fig:gluonloop}
  \end{center} 
\end{figure} 

In the imaginary-time formalism one obtains for the goldstino self-energy 
(cf.~fig.~\ref{fig:gluonloop}) with momentum $P$ summed over helicities:
\beq
  \label{eq:S1}
\Sigma(P) = \mbox{tr}\left[ \sum_{l=\pm 1/2}\psi^l(P)\overline{\psi}^l(P)
\Si_{\gr}(P)\right]
  = T\sum_{k_0} \int \frac{d^3k}{(2\pi)^3} \mbox{tr} \left[
    \slP V_2^\nu {1\over \slQ} \D_{\mu\nu}(K) V_1^\mu\right].
\end{equation}
Here we have neglected gluino and gravitino masses since $\msq,\mgl \ll T$;
$V_{1,2}^\mu$ are the vertices, $Q=P-K$ is the momentum of the gluino, 
and $\D_{\mu\nu}(K)$ is the resummed gluon 
propagator, which is obtained from the resummed photon propagator
(\ref{eq:effgluonprop}) by the substitution 
$m_\g \rightarrow m_g$. The thermal gluon mass for $N$ colours and $n_f$
colour triplet and anti-triplet chiral multiplets is given by  
\beq
  \label{eq:gluonmass}
  m_g^2 = \frac{g^2 T^2}{6}(N + n_f)\;.
\eeq
This result is easily obtained from the expressions for the gluon vacuum 
polarization \cite{Bra90} by adding up the contributions
from gluons, gluinos, quarks and squarks. 

Inserting gluon propagator and
vertices in eq.~(\ref{eq:S1}) yields the gauge-independent result
\beq
  \label{eq:S2}
  \Sigma(P) = \frac{4}{3}\frac{\mgl^2 T}{M^2\mgr^2}(N^2-1) \sum_{k_0} 
  \int\frac{d^3k}{(2\pi)^3} \left(D_L \D_L + D_T \D_T\right) \frac{1}{Q^2}\;,
\eeq
where
\begin{eqnarray}
  \label{eq:traces}
  D_T(k_0,k,E,p,\pk) & = & \frac{1}{32}\mbox{tr}\left\{\slP\ [\slK,\g^\nu]\
  \slQ\ [\slK,\g^\mu] A_{\mu\nu}\right\}\;,\nonumber\\
  D_L(k_0,k,E,p,\pk) & = & \frac{1}{32}\mbox{tr}\left\{\slP\ [\slK,\g^\nu]\
  \slQ\ [\slK,\g^\mu] B_{\mu\nu}\right\}\;.
\end{eqnarray}
After a straightforward calculation, analogous to the one for the axion
self-energy, one finds for the gravitino production rate
\bea
  \label{eq:gluonsoft1}
\Gs(E) &=& 
\left.-\frac{\mbox{Im}{\Sigma(E+i\epsilon,p)}}{E}\right|_{k<k_{\rm cut}}
\nonumber \\
&=& \frac{\mgl^2 (N^2-1)T}{6\pi M^2\mgr^2} \int_0^{k_{\rm cut}} dk k^3 
\int_{-k}^k {d\om\over \om} \nonumber\\
&&\hspace{2cm}\times\left[\rho_L(\om,k)\left(1-{\om^2\over k^2}\right) 
+ \rho_T(\om,k)\left(1-{\om^2\over k^2}\right)^2 \right]\; .
\eea
The momentum integral depends logarithmically on the cutoff 
$k_{\rm cut}$. The integrand, which is identical with the one for the axion
production rate (\ref{eq:gammasoft1}), agrees with the result obtained in 
\cite{Ell96}. After performing the momentum integrations one finally obtains
\beq
\label{eq:gluonsoft2}
  \Gs(E) = \frac{(N^2-1)\mgl^2 m_g^2 T}{4\pi M^2 \mgr^2}
  \left[\ln\left(\frac{k_{\rm cut}^2}{m_g^2}\right)-1.379\right].
\eeq
Note, that the overall normalization differs from the expression given in 
\cite{Ell96} by the factor $4(N^2-1)$.

The dependence of the soft part of the gravitino production rate on the
cutoff $k_{\rm cut}$ is again cancelled by the cutoff dependence of the
contribution from the hard $\ttt$ processes. There are 10 processes
denoted by A to J \cite{Ell84}:

\begin{itemize}
  \item A: $g^a + g^b \rightarrow \gl^c + \gr$ 
\begin{center}
  \epsfig{file=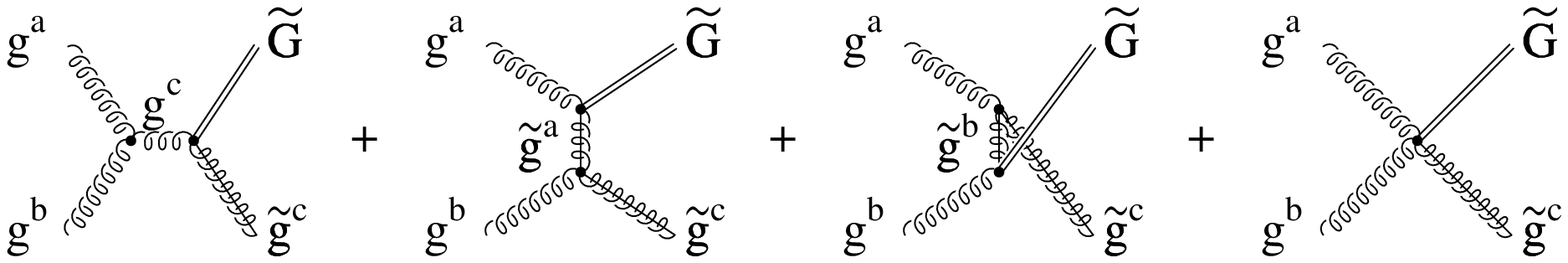,width=30em}
\end{center}
  \item B: $g^a + \gl^b \rightarrow g^c + \gr$ (crossing of A)
  \item C: $\sq_i + g^a \rightarrow \sq_j + \gr$
\begin{center}
  \epsfig{file=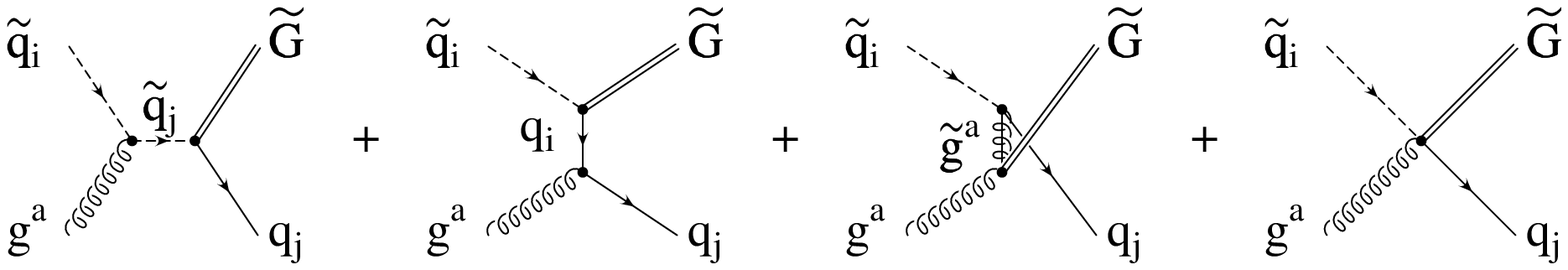,width=30em}
\end{center}
  \item D: $g^a + q_i \rightarrow \sq_j + \gr$ (crossing of C)
  \item E: $\bar{\sq_i} + q_j \rightarrow g^a + \gr$ (crossing of C)
  \item F: $\gl^a + \gl^b \rightarrow \gl^c + \gr$
\begin{center}
  \epsfig{file=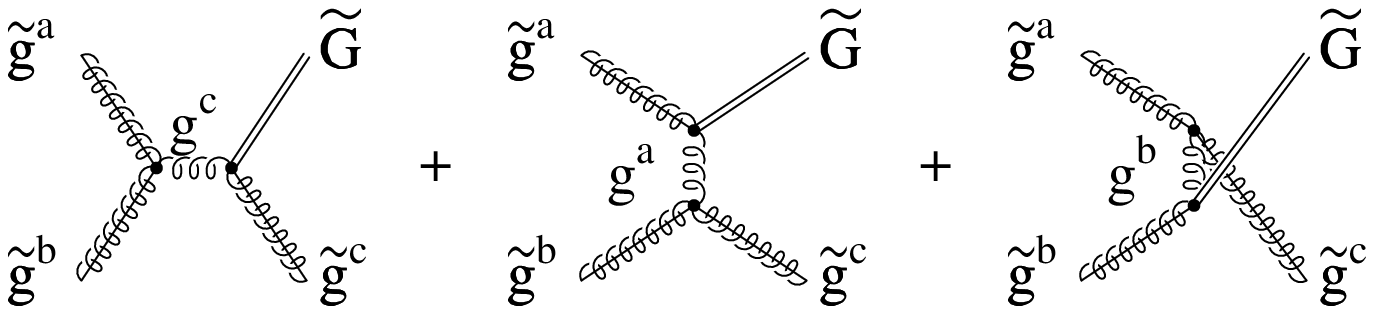,width=24em}
\end{center}
  \item G: $q_i + \gl^a \rightarrow q_j + \gr$
\begin{center}
  \epsfig{file=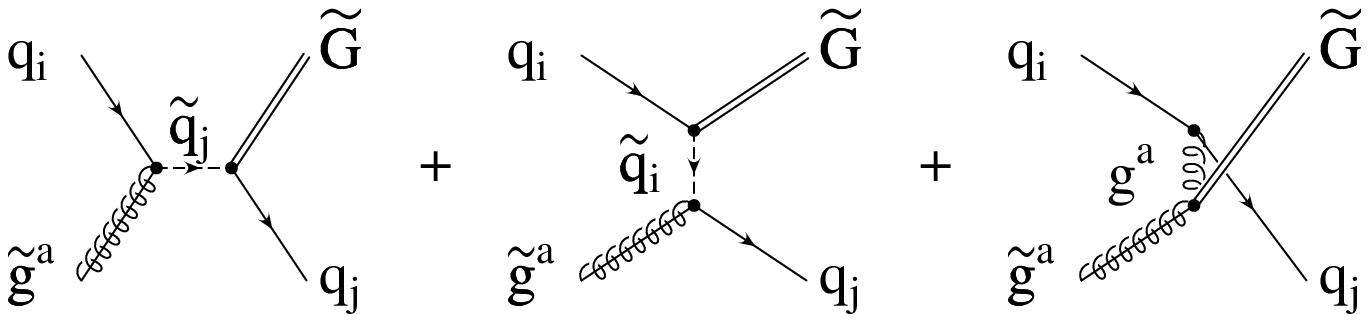,width=24em}
\end{center}
  \item H: $\sq_i + \gl^a \rightarrow \sq_j + \gr$
\begin{center}

  \epsfig{file=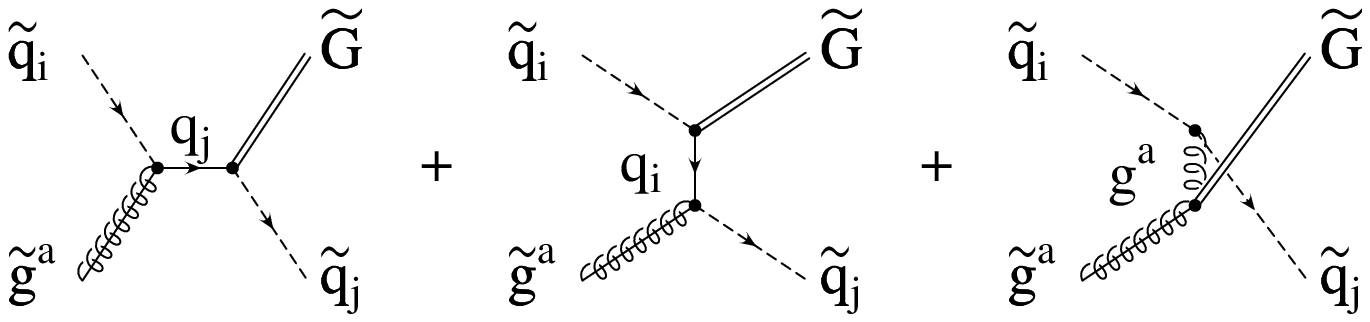,width=24em}
\end{center}
  \item I: $q_i + \bar{q_j} \rightarrow \gl^a + \gr$ (crossing of G)
  \item J: $\sq_i + \bar{\sq_j} \rightarrow \gl^a + \gr$ (crossing of H)
\end{itemize}
The corresponding matrix elements have been evaluated in \cite{Bol98}.
As discussed in section~2, they must have the form   
\beq
|\Mc_i|^2 \propto \frac{1}{M^2}\factor\;
\eeq
in the high energy limit. 

\begin{table}[h]
\begin{center}
\begin{tabular}{r|c|l}
 & process $i$ & 
$|\Mc_i|^2/\frac{g^2}{M^2}\factor$\\ 
\hline \hline
A & $g^a + g^b \rightarrow \gl^c + \gr $ & 
$4(s+2t+2\frac{t^2}{s})|f^{abc}|^2$ \\ \hline 

B & $g^a + \gl^b \rightarrow g^c + \gr $ & 
$-4(t+2s+2\frac{s^2}{t})|f^{abc}|^2$\\ \hline

C & $\sq_i  + g^a \rightarrow q_j + \gr $ & 

$2s|T^a_{ji}|^2$\\ \hline

D & $g^a + q_i \rightarrow \sq_j + \gr $ & 
$-2t|T^a_{ji}|^2$\\ \hline 

E & $\bar{\sq}_i + q_j \rightarrow g^a + \gr $ & 
$-2t|T^a_{ji}|^2$\\ \hline

F & $\gl^a + \gl^b \rightarrow \gl^c + \gr $ & 
$-8\frac{(s^2+st+t^2)^2}{st(s+t)}|f^{abc}|^2$\\ \hline

G & $q_i + \gl^a \rightarrow q_j + \gr $ & 
$-4(s+\frac{s^2}{t})|T^a_{ji}|^2$\\ \hline

H & $\sq_i + \gl^a \rightarrow \sq_j + \gr $ & 
$-2(t+2s+2\frac{s^2}{t})|T^a_{ji}|^2$\\ \hline

I & $q_i + \bar{q}_j \rightarrow \gl^a + \gr $ & 
$-4(t+\frac{t^2}{s})|T^a_{ji}|^2$\\ \hline

J & $\sq_i + \bar{\sq}_j \rightarrow \gl^a + \gr $ & 
$2(s+2t+2\frac{t^2}{s})|T^a_{ji}|^2$

\end{tabular}
\medskip
\caption{\it Squared matrix elements for gravitino ($\gr$)
production in two-body processes involving left-handed quarks ($q_i$), 
squarks ($\sq_i$), gluons ($g^a$) and gluinos ($\gl^a$). The values are
given for the specified choice of colors and summed over spins in the
initial and final state. $f^{abc}$ and $T^a_{ji}$ 
are the usual SU(3) colour matrices. }
\label{tab:diffcs}
\end{center}
\end{table}

In table~\ref{tab:diffcs} 
the squared matrix elements of all ten processes are listed. Sums over initial
and final spins have been performed. For quarks and squarks the contribution
of a single chirality is given. One easily checks that the matrix elements
satisfy the relevant crossing symmetries. The particle momenta $P_1$,
$P_2$, $P_3$, and $P$ used in the calculations correspond 
to the particles  in the order in which they are written down in the
column "process $i$"   of table~\ref{tab:diffcs}. This fixes the energies of
Bose and Fermi distribution. The matrix elements in the table correspond
to the definitions $s=(P_1 + P_2)^2$ and $t=(P_1 - P_3)^2$.

The different processes fall into three classes depending on the number of
bosons and fermions in initial and final state. A, C and J are BBF processes
with two bosons in the initial and a fermion in the final state; 
correspondingly, B, D, E and H are BFB processes, and F, G and I are FFF
processes. Only four processes, B, F, G and H contribute to the logarithmic
cutoff dependence. The gravitino production rate is then given by 
(cf.~\cite{Wel83}),
\bea
\label{eq:hardgluon}
&&n_F(E)\G_{\gr}^{\rm hard}(E) 
= \int {d\Om_p\over 4\pi}{1\over 2E}\int\left[\prod_{i=1}^3
\frac{d^3p_i}{(2\pi)^3 2E_i}\right](2\pi)^4\delta^4(P_1+P_2-P-P_3)\nonumber \\
&&\hspace{0.3cm} \times \left(n_{BBF}\ |M_{BBF}|^2 + n_{BFB}\ |M_{BFB}|^2 
+ n_{FFF}\ |M_{FFF}|^2\right)\Theta(|{\bf p}_1-{\bf p}_3|-k_{\rm cut})\ .
\eea
Here, $n_{BBF}$, $n_{BFB}$ and $n_{FFF}$ are the products of number densities 
for the corresponding processes, e.g.,
\beq
n_{BBF} =  n_B(E_1) n_B(E_2) (1 - n_F(E_3))\;.
\eeq

The matrix elements $|M_{BBF}|^2$ etc. are obtained by summing the 
corresponding matrix elements
in table~\ref{tab:diffcs} with the appropriate multiplicities and statistical
factors. Angular and momentum integrations can now be carried out as in the
case of axion production. One finally obtains the result 
\bea\label{gluonhard}
\G_{\gr}^{\rm hard}(E)&=&\factor\frac{g^2 (N^2-1)}{8\pi^3 M^2}
\Bigg\{2\pi^2(N+n_f)T^3 \left(
\ln\left(\frac{2T}{k_{\rm cut}}\right)+\frac{17}{6}-\gamma+\frac{\zeta'(2)}
{\zeta(2)}\right)  \nonumber \\
&&+ (N+n_f) \left(e^{E/T}+1\right)
\int_0^\infty dE_3 \int_0^{E+E_3}dE_1 \ln\left(\frac{|E_1-E_3|}{E_3}\right)
\nonumber \\ 
&&\hspace{1cm}\times
\Bigg[\Theta(E-E_1)\frac{d}{dE_1}\left[(n_{BFB}+n_{FFF})
(\frac{E_1^2E_2^2}{E^2}-E_3^2)\right]\nonumber \\
&&\hspace{1.3cm} -\Theta(E_1-E_3)
\frac{d}{dE_1}\left[(n_{BFB}+n_{FFF})\frac{E_2^2}{E^2}(E_1^2+E_3^2)\right]\nonumber\\
&&\hspace{1.3cm} + \Theta(E_3-E_1)
\frac{d}{dE_1}\left[(n_{BFB}+n_{FFF})(E_1^2+E_3^2)\right]\Bigg]
\nonumber\\
&&\hspace{0.5cm} + I_{BBF} + I_{BFB} + I_{FFF}\Bigg\}\;.
\eea
Performing the differentiations with respect to $E_1$ yields expressions
analogous to the one given in eq.~(\ref{chard}). $I_{BBF}$, $I_{BFB}$ and 
$I_{FFF}$, which are not all proportional to $N+n_f$, are given in 
appendix~C; they 
contribute to the cutoff-independent part of $\G_{\gr}^{\rm hard}$. 

The dependence on $k_{\rm cut}$ cancels in the sum of soft and hard 
contributions to the production rate. From eqs.~(\ref{eq:gluonsoft2}) and 
(\ref{gluonhard}) one obtains for the collision term
\bea\label{eq:collgrav}
C_{\gr}(T) &=& \int {d^3p\over (2\pi)^3} n_F(E) 
\left(\Gs(E)+\Gamma_{\gr}^{\rm hard}(E)\right) \nonumber\\
&=& \factor \frac{3\zeta(3) g^2 (N^2-1)T^6}{32\pi^3 M^2}\nonumber\\
&& \hspace{1.5cm}
\Bigg\{\left[\ln\left(\frac{T^2}{m_g^2}\right) + 0.3224\right] (N+n_f) +
0.5781 n_f \Bigg\}.
\eea
This is the main result of this paper.
It allows to calculate the gravitino abundance to leading order in the 
gauge coupling $g(T)$, contrary to previous estimates which depended
either on ad hoc cutoffs \cite{Ell84,Mor93} or on an unknown scale of the
logarithmic term \cite{Bol98}.

An important question concerns the size of higher-order corrections. Note, that
$g\simeq 0.85$ for $T \sim 10^{10}$~GeV. This is much better than at the 
electroweak scale $T\sim 100$ GeV where $g\simeq 1.2$,
or for the quark-gluon plasma at $T\sim 1$ GeV where
$g\simeq 2.5$. However, one still has to worry about 
the usually assumed separation of scales
$g^2 T \ll gT \ll T$, which would correspond to $\mu_g\ll m_g\ll T$, where
$\mu_g\sim g^2T$ is the magnetic screening mass.  
Note, that for the supersymmetric standard model with 
$N=3$ and $n_f=6$ one has $m_g\simeq T$. For the static 
Debye and magnetic screening masses
the separation of scales has recently been studied in detail
for the case of non-supersymmetric QCD \cite{HaLaPh00}.
For real-time processes almost nothing is presently 
known about non-perturbative effects 
related to the magnetic screening mass. 
This is a challenging theoretical
problem.
\section{Gravitinos as cold dark matter}
\label{chap:scenario}
\ 
\vskip -0.5cm
We can now study the cosmological implications of our result 
eq.~(\ref{eq:collgrav}) for the Boltzmann collision term of gravitino 
production. We are particularly interested in the case of large reheating
temperatures after inflation, i.e. $T_R \simeq 10^8 - 10^{10}$~GeV, which are 
relevant for models of leptogenesis. In the following we shall concentrate
on the possibility that the gravitino is the lightest supersymmetric
particle (LSP), updating the discussion in \cite{Bol98}, where it was pointed 
out that a large gravitino mass $m_{\tilde{G}}\sim 100$ GeV 
is compatible with such reheating temperatures. We shall ignore 
the non-thermal production of gravitinos \cite{Kal00,Giu99} which
depends on the model of inflation. 

From the Boltzmann equation, 
\beq
    \frac{d\ngr}{dt} + 3 H \ngr = C_{\gr}\;,
    \label{eq:Beq}
\eeq
one obtains for the gravitino abundance at temperatures $T<T_R$, assuming 
constant entropy, 
\beq
  \label{eq:ygra1}
  \Ygr(T) = \frac{\ngr(T)}{\nrad(T)} \simeq 
  \frac{g_{\star S}(T)}{g_{\star S}(T_R)}
    \frac{C_{\gr}(T_R)}{H(T_R)\nrad(T_R)}\;,
\eeq
where $g_{\star S}(T)$ is the number of effectively massless degrees
of freedom \cite{RPP00}. For $T<1$~MeV, i.e. after nucleosynthesis, 
$g_{\star S}(T)=\frac{43}{11}$, whereas $g_{\star S}(T_R)=\frac{915}{4}$ 
in the supersymmetric standard model.
With $ H(T) = (g_\star(T)\pi^2/90)^{1/2} T^2/M$ 
one obtains in the case of light gravitinos 
($\mgr\ll \mgl(\mu)$, $\mu\simeq 100$~GeV)
from eqs.~(\ref{eq:ygra1}) and (\ref{eq:collgrav})
for the gravitino abundance and for the contribution to $\Om h^2$, 
\begin{equation}
    \Ygr = 1.1\cdot 10^{-10}
    \left(\frac{T_R}{10^{10}\,\mbox{GeV}}\right)
    \left(\frac{100\,\mbox{GeV}}{m_{\gr}}\right)^2
    \left(\frac{\mgl(\mu)}{1\,\mbox{TeV}}\right)^2,
    \label{eq:ygra2}
\end{equation}
\begin{eqnarray}
    \Ogr h^2 & = & \mgr \Ygr(T) \nrad(T) h^2 \rho_c^{-1} \nonumber \\
    & = & 0.21
    \left(\frac{T_R}{10^{10}\,\mbox{GeV}}\right)
    \left(\frac{100\,\mbox{GeV}}{\mgr}\right)
    \left(\frac{\mgl(\mu)}{1\,\mbox{TeV}}\right)^2.    
\label{eq:omgr1}
\end{eqnarray}
Here we have used $g(T_R)=0.85$, $\nrad(T)= \zeta(3)T^3/\pi^2$, 
and $\mgl(T)= g^2(T)/g^2(\mu) \mgl(\mu)$; 
$\rho_c=3H_0^2M^2=1.05 h^2 10^{-5}$\,GeV\,cm$^{-3}$ is the critical 
energy density. 
 \begin{figure}[h]
  \begin{center}
  \epsfig{file=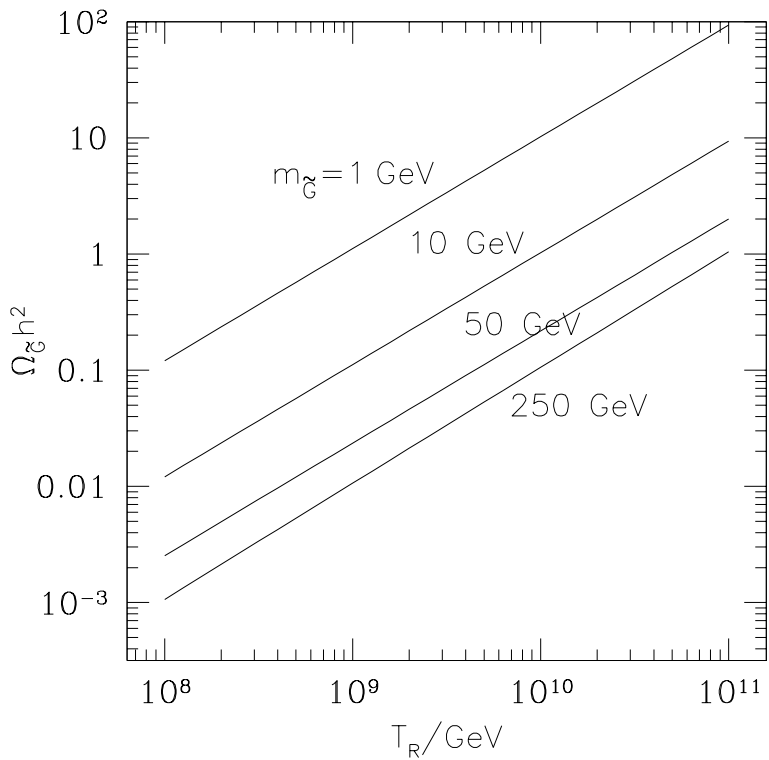,width=25em}
  \caption{\it The density parameter $\Ogr h^2$
for different gravitino masses $\mgr$ as function of the reheating
temperature $T_R$. The gluino mass has been set to $\mgl=700$~GeV.}
  \label{fig:omega}
  \end{center}  
\end{figure}
The new result for $\Ogr h^2$ is smaller by a factor of 3 compared to the
result given in \cite{Bol98}. Due to the large value of the plasma mass 
$m_g$ an estimate of the gravitino production rate, which is based 
just on the logarithmic term of the $2\to 2$ cross sections 
as in \cite{Bol98} is rather uncertain.

It is remarkable that reheating temperatures 
$T_R \simeq 10^8 - 10^{10}$~GeV lead to values 
$\Ogr h^2=0.01\dots 1$ in an interesting gravitino mass range. 
This is illustrated in fig.~\ref{fig:omega} for a gluino mass 
$\mgl = 700$~GeV. As an example, for $T_R\simeq 10^{10}$~GeV, 
$\mgr \simeq 80$~GeV and $h \simeq 0.65$ \cite{RPP00} one finds 
$\Ogr=0.35$, which agrees with
recent measurements of $\Om_{\rm M}$ \cite{RPP00}. 
\begin{figure}[h]
  \begin{center}
  \epsfig{file=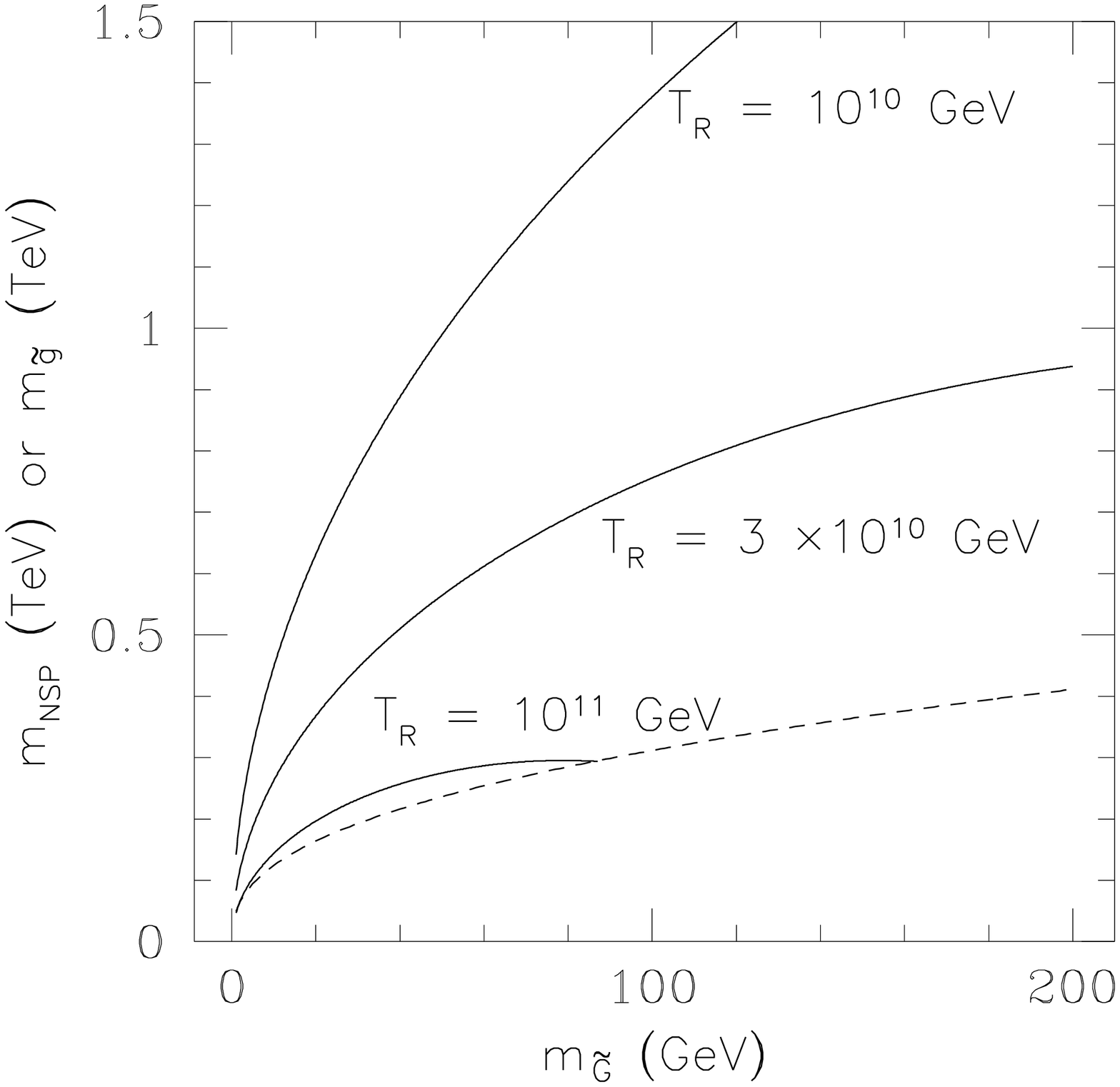,width=25em}
\caption{\it Upper and lower bounds on the gluino mass and the NSP mass
as functions of the gravitino mass. The full lines represent the upper bound 
on the gluino mass $\mgl > m_{NSP}$ for different reheating 
temperatures from the closure limit constraint. 
The dashed line is the lower bound on $m_{NSP}$ which follows
from the NSP lifetime.}
  \label{fig:masses}
  \end{center}  
\end{figure}
In general, to find a viable cosmological scenario one has to avoid two 
types of gravitino problems: For unstable gravitinos their decay products
must not alter the observed abundances of light elements in the universe, 
which is referred to as the big bang nucleosynthesis (BBN) constraint.
For stable gravitinos this condition has to be met by other super particles, 
in particular the next-to-lightest super particle (NSP), which decay into
gravitinos; further, the contribution of gravitinos 
to the energy density of the universe
must not exceed the closure limit, i.e. 
$\Ogr = \rho_{\gr}/\rho_c < 1$.
Consider first the constraint from the closure limit. 
The condition $\Ogr = \Ygr \mgr \nrad/\rho_c \leq 1$
yields an allowed region in the $\mgr$-$\mgl$ plane which is shown
in fig.~\ref{fig:masses} for three different values of the reheating
temperature  $T_R$. The allowed regions are below the solid lines, 
respectively.

With respect to the BBN constraint, consider a
nonrelativistic particle $X$ decaying into electromagnetically and strongly
interacting relativistic particles with a lifetime $\tau_X$. X decays change
the abundances of light elements the more
the longer the lifetime $\tau_X$ and the higher the energy density 
$m_X Y_X \nrad$ are. These constraints have been studied in detail 
by several groups 
\cite{Ell92,Kaw95,Hol99}. They rule out the possibility of unstable gravitinos
with $\mgr \sim 100$~GeV for $T_R \sim 10^{10}$~GeV.

For stable gravitinos the NSP plays the role 
of the particle X. The lifetime of a fermion decaying into its scalar partner 
and a gravitino is 
\beq
\tau_\mathrm{NSP} = 48\pi \frac{\mgr^2 M^2}{m_\mathrm{NSP}^5}\ .
\eeq
For a sufficiently short lifetime, 
$\tau_\mathrm{NSP}<2\cdot 10^6\,\mbox{s}$, the energy density which becomes
free in NSP decays is bounded by $m_X Y_X < 4\cdot 10^{-10}\,\mbox{GeV}$, 
which corresponds to $\Om_X h^2 < 0.008$. The lifetime constraint yields
a lower bound on super particle masses which is represented by the dashed
line in the $\mgr$-$m_{\mathrm{NSP/\gl}}$ plane in  Fig.~\ref{fig:masses}. 
 
In order to decide whether the second part of the BBN constraint, 
$\ONSP h^2 < 0.008$,  is satisfied, one has  to 
specify which particle is the NSP. The case of a higgsino-like neutralino  
as NSP  has been discussed in \cite{Bol98}. A detailed discussion of the
case where a scalar $\tau$-lepton is the NSP has been given in 
\cite{Ger99},\cite{Asa00}. 

A complete treatment of gravitinos as cold dark matter has to 
include non-thermal contributions. The situation is analogous to leptogenesis 
where, in principle, non-thermal contributions also have to be added
to the thermal part.
However, non-thermal contributions depend on assumptions about the state 
of the early universe before the hot thermal phase, for instance the type of
inflationary phase, and they are therefore strongly model dependent. 

\section{Outlook}
\ 
\vskip -0.5cm
The main result of the paper is the production rate of gravitinos for
supersymmetric QCD at high temperature to leading order in the gauge
coupling. The result is valid for gravitino masses larger or smaller
than the gluino mass.

As expected the gravitino production rate depends logarithmically on the
gluon plasma mass which regularizes an infrared divergence occuring in leading
order. Following the procedure of Braaten and Yuan, the result is
obtained by matching contributions to the gravitino self-energy with
soft and hard internal gluon momenta and by using a resummed gluon propagator
for the soft part. As a byproduct a new result for the axion production
rate in a QED plasma is obtained which is slightly smaller than a
previously published result.

The QCD coupling is large, and even at temperatures $T\sim 10^{10}$ GeV 
the usually assumed separation of scales, $g^2 T \ll g T \ll T$ appears
problematic. 
Hence, higher-order corrections to the gravitino production rate
may be sizeable. Further, it is of crucial importance to gain some 
understanding of the influence of the non-perturbative magnetic mass scale
$\mu_g$ on real-time processes in general.

The thermal gravitino production rate plays a central role in cosmology
since it is closely related to the dark matter problem. For many 
supersymmetric extensions of the standard model this rate defines a
limiting temperature beyond which the standard hot big bang picture becomes
inconsistent. At present supersymmetric theories offer several interesting
candidates for cold or hot dark matter. It is an intriguing possibility
that the gravitino itself is the dominant component of cold dark matter.
\\
\vskip 0.1cm
\noindent
We would like to thank T.~Asaka, O.~B\"ar, D.~B\"odeker, O.~Philipsen
and M.~Pl\"umacher for helpful discussions. The work of A.B. has been supported
by a Heisenberg grant of the D.F.G. 

\newpage
\section*{Appendix A}
\ 
\vskip -0.5cm
\label{chap:appendix}
\setcounter{equation}{0}
\renewcommand{\theequation}{A.\arabic{equation}}

In the following we shall derive the prefactor of the self-energy,
\beq\label{prove}
\Sigma(P) \propto \factor\;,
\eeq 
extending the discussion in section~2.

\begin{figure}[h]
  \begin{center}
  \epsfig{file=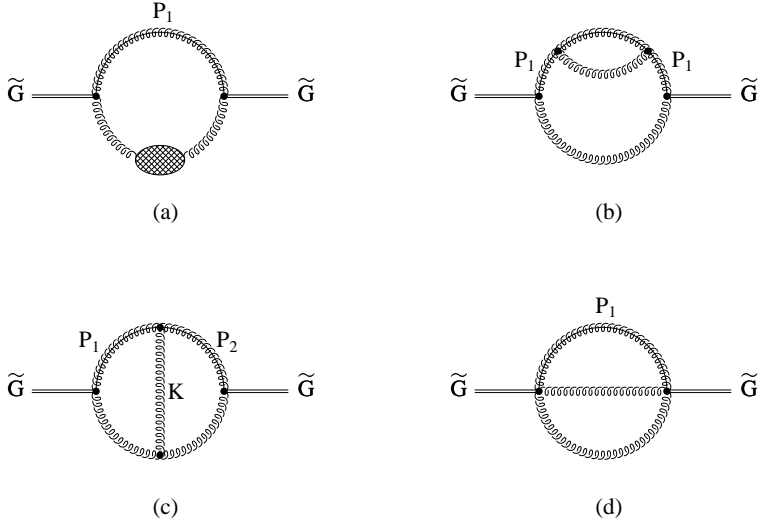,width=25em}
  \caption{\it Contributions to the gravitino self energy. }
  \label{fig:master_loop}
  \end{center} 
\end{figure} 

The gravitino self-energy takes the form (cf.~eq.~(\ref{decomp1})),
\bea
\Sigma(P) 
&\propto&  {1\over M^2} \mbox{tr}{\left[\Pmn S(P)^\nu \ldots
\overline{S}^\mu(P)\right]}\nonumber\\ 
&\propto& {1\over M^2} \mbox{tr}\left[
(- \slP)[\g^\nu,\g^\rho]\g^\mu \lambda^a(P_1)F^a_{\nu\rho}(K_1) \ldots 
F^b_{\tau\sigma}(K_1) \overline{\lambda}^b(P_1) \g_\mu [\g^\tau,\g^\sigma]
\right] \nonumber\\
&& +  {2 m_{\gl}^2\over 3 m_{\gr}^2 M^2} \mbox{tr}\left[
[\g^\nu,\g^\rho] \lambda^a(P_1)F_{\nu\rho}(K_1) \ldots
F_{\tau\sigma}(K_1) \overline{\lambda}^a(P_1) [\g^\tau,\g^\sigma]\right] \;.
\eea
The one- and two-loop contributions for the pure gauge theory in resummed 
perturbation theory are depicted in Fig.~(\ref{fig:master_loop}). The 
contributions (a) and (b) represent for the gluino line  the first two 
terms of the gluino propagator. This corresponds to the substitution,
\beq
\lambda(P_1)\ldots \overline{\lambda}(P_1) \rightarrow
\slP_1 A(P_1,v) + \slv B(P_1,v)\;.
\eeq
This is the general form of the gluino propagator for $\mgl = 0$ because of
chiral symmetry and the fact that the velocity $v$ appearing in the gluon
propagator is the only other vector available apart from the momentum $P_1$.
With 
\beq
\g^\mu \g^\nu \g_\mu = -2 \g^\nu
\eeq
one then reads off the factor (\ref{prove}) for the contributions (a) and
(b). The same arguments apply for the contribution Fig.~(6d).

For Fig.~(6c) one obtains for the gluino line,
\beq
\lambda(P_1)\ldots \overline{\lambda}(P_1) \rightarrow
\slP_1 \g^\nu \slP_2\; .
\eeq
With
\beq
\g^\mu \slP_1 \g^\nu \slP_2 \g_\mu = - 2 \slP_2 \g^\nu \slP_1 \;,
\eeq
the interchange $P_1 \leftrightarrow P_2$ and the property of the gluon
propagator $\Delta(-K) = \Delta(K)$ one obtains the factor (\ref{prove})
also for this contribution.

\section*{Appendix B}
\label{chap:appendixb}
\ 
\vskip -0.5cm
\setcounter{equation}{0}
\renewcommand{\theequation}{B.\arabic{equation}}
In this appendix we explain the calculation of the 
contribution of hard virtual photons to the axion
production rate $\G_a^{\rm hard}(E)$. 
We start by reconsidering the defining 
equation (\ref{eq:harddef}):
 \bea \label{eq_a:harddef}
n_B(E)\G_a^{\rm hard}(E) 
&=& 2\int {d\Om_p\over 4\pi}{1\over 2E}\int\left[\prod_{i=1}^3
\frac{d^3p_i}{(2\pi)^3 2E_i}\right](2\pi)^4\delta^4(P_1+P_2-P-P_3)\nonumber \\
&& \ \ n_{\rm total}^{FBF}
|M|^2\Theta(|{\bf p}_1-{\bf p}_3|-k_{\rm cut})\;
\eea
where the matrix element squared for $e^-\gamma\to e^-a$ is given 
in Eq.~(\ref{eq:axionmat}) and
\bea
n_{\rm total}^{FBF}= n_F(E_1)n_B(E_2)(1 - n_F(E_3)).
\eea 
A great simplification is achieved in the
computation of (\ref{eq_a:harddef}) if one uses as reference
momentum the difference vector ${\bf k} ={\bf p}_1-{\bf p}_3$, i.e.,
we write
\begin{eqnarray}
\frac{d^3p_1}{2E_1}&=&\delta(P_1^2)\Theta(E_1)dE_1d^3p_1\nonumber \\
&=& \int d^3k \delta^3({\bf k} +{\bf p}_3-{\bf p}_1)
\delta(P_1^2)\Theta(E_1)dE_1d^3p_1\nonumber \\
&=& \delta(E_1^2-|{\bf k}+{\bf p}_3|^2)\Theta(E_1)dE_1d^3k.
\end{eqnarray}
Further,
\begin{eqnarray}
\frac{d^3p_2}{2E_2}\delta^4(P_1+P_2-P-P_3)&=&
\delta(P_2^2)\Theta(E_2)d^4P_2\delta^4(P_1+P_2-P-P_3) \nonumber \\
&=& \delta((E+E_3-E_1)^2-({\bf p}-{\bf k})^2)\Theta(E+E_3-E_1).
\end{eqnarray}
We now use rotational invariance to choose
\begin{eqnarray}
{\bf k} &=& k\ (0,0,1),\nonumber \\ 
{\bf p} &=& E\ (0,\sin\tilde{\theta},\cos\tilde{\theta}),\nonumber \\
{\bf p}_3 &=& E_3\ (\cos\phi\sin\theta,\sin\phi\sin\theta,\cos\theta),
\end{eqnarray}
which implies
\begin{eqnarray}
s&=& (P_1+P_2)^2=(P+P_3)^2=2EE_3(1-\sin\theta\sin\phi\sin\tilde{\theta}
-\cos\theta\cos\tilde{\theta}),\nonumber \\
t&=& (P_1-P_3)^2=(E_1-E_3)^2-k^2,
\end{eqnarray}
and
\begin{eqnarray}
|{\bf k}+{\bf p}_3|^2&=&E_3^2+k^2+2E_3k\cos\theta,\nonumber \\
|{\bf p}-{\bf k}|^2&=&E^2+k^2-2Ek\cos\tilde\theta.
\end{eqnarray}
It follows that
\begin{eqnarray}
 \delta((E+E_3-E_1)^2-({\bf p}-{\bf k})^2)&=&
\frac{1}{2kE}
\delta\left(\cos\tilde\theta-\frac{E^2+k^2-(E+E_3-E_1)^2}
{2kE}\right),\nonumber \\
\delta(E_1^2-|{\bf k}+{\bf p}_3|^2)&=&
\frac{1}{2kE_3}
\delta\left(\cos\theta-\frac{E_1^2-E_3^2-k^2}{2kE_3}\right).
\end{eqnarray}
The integrations over the $\delta$-functions yield the following 
$\Theta$-functions (where we use also the $\Theta$-functions
$\Theta(E_1),\Theta(E+E_3-E_1)$ and $E= p>0,E_3=p_3>0 $:\\
1.) From the integration over $\cos\theta$ we get:
\begin{eqnarray}
\cos\theta<1  &\rightarrow& k > E_1 - E_3,\nonumber \\
\cos\theta>-1 &\rightarrow& E_1 > |E_3-k|.
\end{eqnarray}
The second of these constraints is equivalent to
\begin{eqnarray}
E_3-E_1< k < E_1+E_3.
\end{eqnarray}
2.) From the integration over $\cos\tilde\theta$ we get:
\begin{eqnarray}
\cos\tilde\theta<1  &\rightarrow& |E-k|<E+E_3-E_1,\nonumber \\
\cos\tilde\theta>-1 &\rightarrow& k>E_3-E_1.
\end{eqnarray}
The first of these constraints is equivalent to
\begin{eqnarray}
E_1-E_3< k < 2E+E_3-E_1.
\end{eqnarray}
After integrating out the $\delta$-functions we therefore have:
 \bea \label{eq_a:hardphi}
n_B(E)\G_a^{\rm hard}(E) = \frac{1}{2^7\pi^4}\frac{1}{E^2}
\int dE_1 dE_3 n_{\rm total}^{FBF} dkd\phi
|M|^2\Om,
\eea
where $\Om$ is the product of all $\Theta$-functions that restrict
the integrations over $E_1,E_3$ and $k$,
\bea\label{eq_a:omega}
\Omega &=& \Theta(k-k_{\rm cut})\Theta(k-|E_1-E_3|)\nonumber \\
&& \Theta(E_1+E_3- k)\Theta(2E+E_3-E_1- k)\nonumber\\
&&\Theta(E_1)\Theta(E_3)\Theta(E+E_3-E_1)\;.
\eea
Since only $s$ depends on $\phi$, we can integrate out also this angle
without difficulty:
\bea
\int d\phi |M|^2 &=& 
 \frac{e^2}{f^2}\int d\phi
\left(\frac{-2s^2}{t}-2s-t\right) \nonumber \\ &=&
\frac{3e^2\pi}{2f^2}\left[(E_1-E_3)^2-k^2\right] 
\left(
-1+\frac{2}{3}\frac{E_1^2+E_3^2+2EE_2}{k^2}
\right.\nonumber \\
&&\left.-\frac{(E_3+E_1)^2(E+E_2)^2}{k^4}
\right)
\equiv g.
\eea
We thereby  obtain the result given in Eq.~(\ref{eq:phiint}) in the
main text.
We now rewrite
the expression for (\ref{eq_a:omega}) using
\begin{eqnarray}
\Theta(E_1+E_3-k) = 1 - \Theta(k-E_1-E_3).
\end{eqnarray}
We use $\Theta(k-E_1-E_3)\Theta(k-|E_1-E_3|)
=\Theta(k-E_1-E_3)$ and thus get
\begin{eqnarray}
\label{eq:newomega}
\Omega &=& \Big[\Theta(k-k_{\rm cut})
\Theta(k-|E_1-E_3|)\Theta(2E+E_3-E_1-k)\nonumber \\
&-&\Theta(k-k_{\rm cut}) 
\Theta(k-E_1-E_3)\Theta(2E+E_3-E_1-k)
\Big]\nonumber \\ &\times&
\Theta(E_1)\Theta(E_3)\Theta(E+E_3-E_1).
\end{eqnarray}
We multiply the second term in the brackets of Eq.~(\ref{eq:newomega})
with 1:
\begin{eqnarray}\label{eq:spurious}
1= \Theta(k_{\rm cut}- E_1-E_3)+\Theta(E_1+E_3-k_{\rm cut}),
 \end{eqnarray}
and note that
\bea
&&\Theta(k-k_{\rm cut}) 
\Theta(k-E_1-E_3)\Theta(k_{\rm cut}-E_1-E_3)\nonumber \\=&&
\Theta(k-k_{\rm cut})\Theta(k_{\rm cut}-E_1-E_3),
\eea
and
\bea
&&\Theta(k-k_{\rm cut}) 
\Theta(k-E_1-E_3)\Theta(E_1+E_3-k_{\rm cut})\nonumber \\=&&
\Theta(k-E_1-E_3)\Theta(E_1+E_3-k_{\rm cut}).
\eea
The contribution from the first term on the r.h.s. of Eq.~(\ref{eq:spurious}) 
is zero in the limit
$k_{\rm cut}\to 0$. We see this by integrating over $k$
from $k_{\rm cut}$ to $2E+E_3-E_1$. The resulting expression
has terms $\sim 1/k_{\rm cut}^3$, $\sim 1/k_{\rm cut}^1$. Since
from  $k_{\rm cut}> E_1+E_3$ it follows that both $E_1$ and $E_3$ are smaller
than  $k_{\rm cut}$ it is easy to  see by power counting that the expression
after the $k$ integration is of order $k_{\rm cut}$.
Then we are left to consider:
\begin{eqnarray}
n_B(E)\G_a^{\rm hard}(E) = g_1+g_2,
\end{eqnarray}
where
\begin{eqnarray}
g_1 &=&  \frac{1}{2^7\pi^4}\frac{1}{E^2}\int_0^{\infty}dE_3\int_0^{\infty}dE_1 n_{\rm total}^{FBF}\Theta(E+E_3-E_1)\nonumber \\
&\times&\int
dk\Theta(k-k_{\rm cut})
\Theta(k-|E_1-E_3|)\Theta(2E+E_3-E_1-k)g, \nonumber \\
g_2  &=& - \frac{1}{2^7\pi^4}\frac{1}{E^2}
\int_0^{\infty}dE_3\int_0^{\infty}dE_1 n_{\rm total}^{FBF}
\Theta(E+E_3-E_1)\Theta(E_1+E_3-k_{\rm cut}) \nonumber \\ &\times& \int dk\Theta(k-E_1-E_3)\Theta(2E+E_3-E_1-k)g,
\end{eqnarray}
The integral  over $k$ in $g_2$ is  nonzero only if
\begin{eqnarray}
E_1+E_3<2E+E_3-E_1 \Leftrightarrow E_1<E.
\end{eqnarray}
In the limit $k_{\rm cut}\to 0$ we therefore get: 
\begin{eqnarray}
g_2  =\frac{e^2}{16\pi^3 f^2}\frac{1}{E^2}\int_0^\infty dE_3 \int_0^{E}dE_1 n_{\rm total}^{FBF} (E_1-E)\left[(E+E_1)E_3+E_1(E-E_1)\right],
\end{eqnarray}
We rewrite this result for later use as follows:
\begin{eqnarray}\label{eq:c2partial}
g_2 &=& \frac{e^2}{16\pi^3 f^2}\frac{1}{E^2}\int_0^{\infty} dE_3 \int_0^{E+E_3}
dE_1\ln\left(\frac{|E_1-E_3|}{E_3}\right)  \nonumber \\
&\times&\Theta(E-E_1)\frac{d}{dE_1}\left[ n_{\rm total}^{FBF}(E_1^2E_2^2-E^2E_3^2)\right].
\nonumber \\
\end{eqnarray}
We now turn towards the computation of $g_1$. We first multiply
by 1:
\begin{eqnarray}
g_1=g_1\Theta(k_{\rm cut}-|E_1-E_3|)+g_1\Theta(|E_1-E_3|-k_{\rm cut})\equiv 
g_{11}+g_{12}
\end{eqnarray} 
Note that
\begin{eqnarray}
&& \Theta(k-k_{\rm cut})\Theta(k-|E_1-E_3|)\Theta(k_{\rm cut}-|E_1-E_3|)\nonumber \\
=&&\Theta(k-k_{\rm cut})
\Theta(k_{\rm cut}-|E_1-E_3|)
\end{eqnarray} 
and
\begin{eqnarray}
&& \Theta(k-k_{\rm cut})\Theta(k-|E_1-E_3|)
\Theta(|E_1-E_3|-k_{\rm cut})\nonumber \\
 =&&\Theta(k-|E_1-E_3|)
\Theta(|E_1-E_3|-k_{\rm cut}).
\end{eqnarray} 
We therefore have
\begin{eqnarray}
g_{11} &=& \frac{1}{2^7\pi^4}\frac{1}{E^2}\int_0^{\infty}dE_3\int_0^{\infty}dE_1 
n_{\rm total}^{FBF}\Theta(E+E_3-E_1)
\Theta(k_{\rm cut}-|E_1-E_3|) \nonumber \\ &\times& \int dk \Theta(k-k_{\rm cut})\Theta(2E+E_3-E_1-k)g
\end{eqnarray}
\begin{eqnarray}
g_{12} &=& \frac{1}{2^7\pi^4}\frac{1}{E^2}\int_0^{\infty}dE_3\int_0^{\infty}dE_1 
n_{\rm total}^{FBF}\Theta(E+E_3-E_1)
\Theta(|E_1-E_3|-k_{\rm cut}) \nonumber \\ &\times& \int dk 
\Theta(k-|E_1-E_3|)\Theta(2E+E_3-E_1-k)g.
\end{eqnarray}
Consider first $g_{11}$. The integration of $g$
over $k$ can be carried out easily. We do not write down the result
explicitly but note that it contains terms $\sim 1/k_{\rm cut}^3$ and
$\sim 1/k_{\rm cut}$. 
The integration over $E_1$ is done next.
From $k_{\rm cut}>|E_1-E_3|$ we get
\begin{eqnarray}
E_3-k_{\rm cut}<E_1<E_3+k_{\rm cut}.
\end{eqnarray} 
In the limit  $ k_{\rm cut}\to 0$ we can therefore set $E_1=E_3$
 in the distribution
functions $n_{\rm total}^{FBF}$ and get
\begin{eqnarray}\label{c1}
g_{11} &=& \frac{e^2}{3\pi^3 f^2} n_B(E)
\int_0^{\infty} dE_3 E_3^2 n_F(E_3)(1 - n_F(E_3))\nonumber \\
&=& \frac{e^2T^3}{18\pi f^2}n_B(E).
\end{eqnarray}
Now we turn towards the computation of $g_{12}$.
We insert 
\begin{eqnarray}
1 = \Theta(E_1-E_3)+\Theta(E_3-E_1).
\end{eqnarray}
Then $g_{12}=g_{121}+g_{122}$ with
\begin{eqnarray}
g_{121} &=& 
\frac{1}{2^7\pi^4}\frac{1}{E^2}\int_0^{\infty}dE_3\int_0^{\infty}dE_1 n_{\rm total}^{FBF}\Theta(E+E_3-E_1)
\Theta(E_1-E_3-k_{\rm cut}) \nonumber \\ &\times& \int dk 
\Theta(k-E_1+E_3)\Theta(2E+E_3-E_1-k)g,\nonumber \\
g_{122} &=& 
\frac{1}{2^7\pi^4}\frac{1}{E^2}\int_0^{\infty}dE_3\int_0^{\infty}dE_1
 n_{\rm total}^{FBF}
\Theta(E_3-E_1-k_{\rm cut}) \nonumber \\ &\times& \int dk 
\Theta(k-E_3+E_1)\Theta(2E+E_3-E_1-k)g.
\end{eqnarray}
The integration over $k$ gives 

\begin{eqnarray}
g_{121} &=& \frac{e^2}{16\pi^3 f^2}\frac{1}{E^2}
\int_0^{\infty}dE_3\int_{E_3+k_{\rm cut}}^{E_3+E}dE_1
 n_{\rm total}^{FBF}
\frac{(E_1^2+E_3^2)E_2^2}{E_1-E_3},
\nonumber \\
g_{122} &=& -\frac{e^2}{16\pi^3 f^2}\frac{1}{E^2}\int_0^{\infty} 
dE_3\int_0^{E_3-k_{\rm cut}}dE_1 n_{\rm total}^{FBF}
\frac{E^2(E_1^2+E_3^2)}{E_1-E_3}.
\end{eqnarray}
The logarithmic dependence on $k_{\rm cut}$ 
is extracted by a partial integration with
$f'(E_1) = 1/(E_1-E_3),\ f(E_1) = \ln\left(|E_1-E_3|/E_3\right)$.
\par
For the surface term we get:
\begin{eqnarray}
g_{\rm surface}&=&
-\frac{e^2}{4\pi^3 f^2}n_B(E)\int_0^\infty dE_3 
\ln\left(\frac{k_{\rm cut}}{E_3}\right)
\frac{E_3^2\exp(E_3/T)}{(\exp(E_3/T)+1)^2}\nonumber \\ &=&
\frac{e^2T^3}{24\pi f^2}n_B(E)\left[
\ln\left(\frac{2T}{k_{\rm cut}}\right)+\frac{3}{2}-\gamma+\frac{\zeta'(2)}
{\zeta(2)}\right] 
\end{eqnarray}
In the remaining term, which is given by $-\int dE_1 f(E_1)g'(E_1)$, 
$k_{\rm cut}$ can be set to zero. Writing $g_{12} = g_{\rm surface}
+g_{\rm partial}$ we obtain: 
\bea
g_{\rm partial}&=&-\frac{e^2}{16\pi^3 f^2}\frac{1}{E^2}\int_0^{\infty}dE_3
\int_0^{\infty}dE_1 \Theta(E+E_3-E_1)
\ln\left(\frac{|E_1-E_3|}{E_3}\right)\nonumber \\ &\times&\Bigg\{
\Theta(E_1-E_3)
\frac{d}{dE_1}\left[ n_{\rm total}^{FBF}E_2^2(E_1^2+E_3^2)\right]
\nonumber \\
&-&\Theta(E_3-E_1)
\frac{d}{dE_1}\left[ n_{\rm total}^{FBF}E^2(E_1^2+E_3^2)\right]\Bigg\}
\eea
Performing the differentiation and combining the results for $g_1$ and $g_2$
leads to the final result Eq.~(\ref{chard}) given in the main text.

\section*{Appendix C}
\label{chap:appendixc}
\ 
\vskip -0.5cm
\setcounter{equation}{0}
\renewcommand{\theequation}{C.\arabic{equation}}
In this appendix we describe in some detail the calculation 
of the hard virtual gluon contribution
and of the other non-singular contributions 
to the gravitino
production rate $\G_{\tilde G}^{\rm hard}(E)$.
We start by considering the defining equation (\ref{eq:hardgluon}).
By summing the corresponding squared matrix elements of table 1
with the appropiate multiplicities and statistical factors, we
get 
\bea
|M_{BBF}|^2 &=& \left(1+\frac{m^2_{\tilde{g}}}{3m^2_{\tilde G}}\right)
\frac{2g^2(N^2-1)}{M^2}\left[\left(s+2t+\frac{2t^2}{s}\right)\left(N+n_f\right)
+2sn_f\right],\\
|M_{BFB}|^2 &=& \left(1+\frac{m^2_{\tilde{g}}}{3m^2_{\tilde G}}\right)
\frac{4g^2(N^2-1)}{M^2}\left[\left(-t-2s-\frac{2s^2}{t}\right)\left(N+n_f\right)-2tn_f\right], \\
|M_{FFF}|^2 &=& \left(1+\frac{m^2_{\tilde{g}}}{3m^2_{\tilde G}}\right)
\frac{4g^2(N^2-1)}{M^2}\left(-t-2s-\frac{s^2}{t}
+\frac{s^2}{t+s}-\frac{t^2}{s}\right)\left(N+n_f\right).
\eea
First we note that since $s=-t-u$ we may write
$s+2t=t-u$ and 
\begin{eqnarray}
-\frac{s^2}{t}+\frac{s^2}{s+t}=-\frac{s^2}{t}-\frac{s^2}{u}.
\end{eqnarray}
The difference $t-u$ and $1/t-1/u$ is odd under exchanging
$P_1$ and $P_2$. If the remaining integrand and the measure is even 
under this transformation, the integral over such terms will be zero.
Therefore in $|M_{BBF}|^2$, the contribution of $s+2t$ will give
zero. Further we may trade $s$ in $|M_{BBF}|^2$ with $-2t$. In 
$|M_{FFF}|^2$, we may 
likewise substitute
\begin{eqnarray}
-\frac{s^2}{t}+\frac{s^2}{s+t}\to -\frac{2s^2}{t}.
\end{eqnarray}
Therefore 
only the following squared matrix elements have to be considered:
\bea
|M_1|^2 &=& -t-2s-\frac{2s^2}{t},\nonumber \\
|M_2|^2 &=& t,\nonumber \\
|M_3|^2 &=& \frac{t^2}{s},
\eea
and we replace the matrix elements in (\ref{eq:hardgluon}) by 
\bea
|M_{BBF}|^2 &\to& \left(1+\frac{m^2_{\tilde{g}}}{3m^2_{\tilde G}}\right)
\frac{4g^2(N^2-1)}{M^2}\left[|M_3|^3\left(N+n_f\right)
-2|M_2|^2n_f\right],\\
|M_{BFB}|^2 &=& \left(1+\frac{m^2_{\tilde{g}}}{3m^2_{\tilde G}}\right)
\frac{4g^2(N^2-1)}{M^2}\left[|M_1|^2\left(N+n_f\right)-2|M_2|^2n_f\right], \\
|M_{FFF}|^2 &\to& \left(1+\frac{m^2_{\tilde{g}}}{3m^2_{\tilde G}}\right)
\frac{4g^2(N^2-1)}{M^2}\left(|M_1|^2
-|M_3|^2\right)\left(N+n_f\right).
\eea
$|M_1|^2$ is the axion matrix element which has been discussed in
appendix B. 
The different statistical factors in the case of gravitino production
do not change the structure of the contribution 
from $|M_1|^2$ as compared to the axion case. Again we can extract
the logarithmic dependence on the cutoff $k_{\rm cut}$ by a partial
integration. In the case of BFB, the surface term contains an
additional term which depends on the energy of the gravitino, 
see Eq.~(\ref{eq:ibfb}) below.
The other two matrix elements do not induce a logarithmic 
dependence on the cutoff $k_{\rm cut}$, i.e. one can set $k_{\rm cut}=0$
to compute their contribution to the gravitino production rate.
The contribution from $|M_2|^2 = t$ can be obtained easily using the same
methods as in the axion case, where now no partial integration is needed.
We obtain
\bea
I_{BBF(BFB)}^t&=&\int {d\Om_p\over 4\pi}{1\over 2E}\int\left[\prod_{i=1}^3
\frac{d^3p_i}{(2\pi)^3 2E_i}\right](2\pi)^4\delta^4(P_1+P_2-P-P_3)
n_{BBF(BFB)}|M_2|^2 \nonumber 
\\ &=& \frac{1}{96\pi^3}
\int_0^{\infty}dE_3\int_0^{E+E_3}dE_1 n_{BBF(BFB)}
\nonumber \\ &\times&
\Bigg\{\Theta(E-E_1)\frac{E-E_1}{E^2}\left[2E^2+(3E_3-E_1)(E+E_1)\right]\nonumber \\
&-&\Theta(E_1-E_3)\frac{E_2^2}{E^2}(2E-E_3+E_1)
\nonumber \\ &+&
\Theta(E_3-E_1)(-3E_3+3E_1-2E)\Bigg\},
\eea
To compute the contribution from  $|M_3|^2 = \frac{t^2}{s}$ it is
convenient to choose different coordinates to perform the angular
integrations, namely
\bea
{\bf q}&\equiv& {\bf p}+{\bf p_3} = q(0,0,1),\nonumber \\
{\bf p} &=& E\ (0,\sin\tilde{\theta},\cos\tilde{\theta}),\nonumber \\
{\bf p}_2 &=& E_2\ (\cos\phi\sin\theta,\sin\phi\sin\theta,\cos\theta).
\eea
The calculation of this contribution to $\G_{\tilde G}^{\rm hard}(E)$
then goes along similar lines as for the axion, i.e. the integration
of the angular variables $\cos\theta, \cos\tilde{\theta}$ can be trivially
performed using the $\delta-$functions. This leads to several constraints
for the integration over $q$, which can be performed without problems.
The final result is rather compact:
\bea
I_{BBF(FFF)}^{t^2/s}
&=&\int {d\Om_p\over 4\pi}{1\over 2E}\int\left[\prod_{i=1}^3
\frac{d^3p_i}{(2\pi)^3 2E_i}\right](2\pi)^4\delta^4(P_1+P_2-P-P_3)
n_{BBF(FFF)}|M_3|^2 \nonumber 
\\ &=& \frac{1}{32\pi^3}
\int_0^{\infty}dE_3\int_0^{E+E_3}dE_2 n_{BBF(FFF)}
\nonumber \\ &\times&
\Bigg\{\frac{E_2^2}{E+E_3}+\Theta(E_2-E_3)\frac{E_3-E_2}{E^2}
\left[E_3(E_3-E_2)+E(E_3+E_2)\right]\Bigg\}.
\eea

The full result for  $\G_{\tilde G}^{\rm hard}(E)$
can be written as in Eq.~(\ref{gluonhard}) with
\bea
I_{BBF}= 32\pi^3\left(e^{E/T}+1\right)\left[(N+n_f)I_{BBF}^{t^2/s}
-2n_fI_{BBF}^t\right],\eea
\bea\label{eq:ibfb}
I_{BFB}= T^3(N+n_f)\left[{\rm Li}_2(-e^{-E/T})-\frac{\pi^2}{6}(1+8\ln(2))\right]-64\pi^3n_f\left(e^{E/T}+1\right)I_{BFB}^t,\eea
\bea
I_{FFF}=-32\pi^3\left(e^{E/T}+1\right)(N+n_f)I_{FFF}^{t^2/s}.
\eea
The hard contribution to the collision term can be obtained
by a numerical integration. Adding all the contributions we finally find:
\bea
C_{\tilde{G}}^{\rm hard}(T) &=& \int\frac{d^3p}{(2\pi)^3}n_F(E)\G_{\tilde G}^{\rm hard}(E) \nonumber \\ &=&\left(1+\frac{m^2_{\tilde{g}}}{3m^2_{\tilde G}}\right)
\frac{3\zeta(3)g^2(N^2-1)T^6}{32\pi^3M^2}\nonumber \\ &\times&
\left\{(N+n_f)
\left[\ln\left(\frac{T^2}{k_{\rm cut}^2}\right)+1.7014\right]
+0.5781 n_f\right\},
\eea
which yields, after adding 
the soft contribution, our final result for the gravitino collision term (\ref{eq:collgrav}).

\newpage

\section*{Erratum}
\ 
\vskip -0.5cm
\setcounter{equation}{0}
\renewcommand{\theequation}{E.\arabic{equation}}

As recently pointed out by Pradler and Steffen (Phys. Rev. {\bf D} 75 (2007)
023509 [hep-ph/0608344]), Eq.~(C.14) has to be replaced by 
\bea
I_{BFB}= -\frac{4}{3}\pi^2\ln(2)T^3(N+n_f) 
-64\pi^3n_f\left(e^{E/T}+1\right)I_{BFB}^t\;.
\eea
The difference between Eqs.~(C.14) and (E.1) is a non-vanishing surface term 
in the BFB case, which in the partial integration leading from (B.24) to 
(B.25) was overlooked. 

As a consequence, the numerical factor multiplying $(N+n_f)$ changes in
Eq.~(C.16) from 1.7014 to 1.8782. Correspondingly, also the numerical
factor multiplying $(N+n_f)$ in Eq.~(44), and the overall factors in 
Eqs.~(47) and (48) change. The corrected equations read
\bea
C_{\gr}(T) &=& \int {d^3p\over (2\pi)^3} n_F(E) 
\left(\Gs(E)+\Gamma_{\gr}^{\rm hard}(E)\right) \nonumber\\
&=& \factor \frac{3\zeta(3) g^2 (N^2-1)T^6}{32\pi^3 M^2}\nonumber\\
&& \hspace{1.5cm}
\Bigg\{\left[\ln\left(\frac{T^2}{m_g^2}\right) + 0.4992\right] (N+n_f) +
0.5781 n_f \Bigg\}\;,
\eea
\begin{equation}
    \Ygr = 1.4\cdot 10^{-10}
    \left(\frac{T_R}{10^{10}\,\mbox{GeV}}\right)
    \left(\frac{100\,\mbox{GeV}}{m_{\gr}}\right)^2
    \left(\frac{\mgl(\mu)}{1\,\mbox{TeV}}\right)^2\;,
\end{equation}
\begin{eqnarray}
    \Ogr h^2 & = & \mgr \Ygr(T) \nrad(T) h^2 \rho_c^{-1} \nonumber \\
    & = & 0.27
    \left(\frac{T_R}{10^{10}\,\mbox{GeV}}\right)
    \left(\frac{100\,\mbox{GeV}}{\mgr}\right)
    \left(\frac{\mgl(\mu)}{1\,\mbox{TeV}}\right)^2\;.    
\end{eqnarray}
Compared to Eqs.~(47) and (48), the gravitino abundance is increased by 30\%.
(Note that also the prefactor $1/(2\sqrt{6})$ in Eq.~(33) has to be replaced
by $1/(4\sqrt{6})$.)\\

\noindent
We thank Josef Pradler and Frank Steffen for informing us about the error
in our paper.

\end{document}